\documentclass{aims}
\usepackage{amsmath}
  \usepackage{paralist}
  \usepackage{graphics} 
  \usepackage{epsfig} 
\usepackage{graphicx}  \usepackage{epstopdf}
 \usepackage[colorlinks=true]{hyperref}
\hypersetup{urlcolor=blue, citecolor=red}

\usepackage[a4paper]{geometry}
\usepackage{amsthm}
\usepackage{amssymb}
\usepackage{amsfonts}
\usepackage{algorithm}
\usepackage{algorithmic}
\usepackage[algo2e]{algorithm2e}
\usepackage{graphicx}
\usepackage{color}
\usepackage[bf,SL,BF]{subfigure}
\usepackage{url}
\usepackage{pdfsync}
\usepackage{blindtext}
\usepackage{epstopdf}
\usepackage{hyperref}
\usepackage{lineno}
\usepackage{tikz}
\usepackage{float}
\usepackage{epsfig,amsbsy,graphicx,multirow}
\usepackage[all]{xypic}

\graphicspath{ {Figures/} }

\def\C{\mathcal{C}}

\def\L{\mathcal{L}}

\def\<{\big\langle}
\def\>{\big\rangle}

\def\phii{\phi_{\mathrm{init}}}
\def\phic{\phi_{\mathrm{cr}}}
\def\phir{\phi_{\mathrm{ref}}}
\def\phim{\phi_{\mathrm{m}}}
\def\Eref{E_{\mathrm{ref}}}

\def\dphi{\delta \phi}
\def\C{\mathcal{C}}
\def\Ci{\C_{\mathrm{init}}}
\def\Cj{\C_{\mathrm{jam}}}
\def\Cr{\C_{\mathrm{ref}}}

\def\Ceq{\C_{\mathrm{eq}}}
\def\wx{\widetilde{x}}

\DeclareMathOperator*{\argmin}{arg\,min}

\definecolor{red}{rgb}{0.9, 0, 0}

\definecolor{lzcol}{rgb}{0.7, 0, 0}
\definecolor{hwcol}{rgb}{0, 0, 0.9}
\definecolor{mlcol}{rgb}{0, 0.7, 0}
\definecolor{todocol}{rgb}{0.0, 0.4, 0.0}

\theoremstyle{remark}
\newtheorem{Remark}{Remark}[section]

\def\currentvolume{X}
 \def\currentissue{X}
  \def\currentyear{200X}
   \def\currentmonth{XX}
    \def\ppages{X--XX}
     \def\DOI{10.3934/xx.xx.xx.xx}

\theoremstyle{definition}

\newtheorem{remark}{Remark}

\title[Energy minimization and preconditioning in Granular Simulation] 
      {Energy Minimization and Preconditioning in the Simulation of Athermal Granular Materials in Two Dimensions}

\author[Haolei Wang and Lei Zhang]{}

\subjclass{Primary: 74G65, 65F08; Secondary: 82B26, 68U20.}
 \keywords{Granular materials, jamming, scaling law, quasi-static evolution, preconditioned energy minimization.}

 \email{hlwang218@sjtu.edu.cn}
 \email{lzhang2012@sjtu.edu.cn}

\thanks{HW and LZ were partially supported by National Natural Science Foundations of China (NSFC 11871339, 11861131, 11571314).}

\thanks{$^*$ Corresponding author: Lei Zhang}

\begin{document}
\maketitle

\centerline{\scshape Haolei Wang and Lei Zhang$^*$}
\medskip
{\footnotesize
 \centerline{School of Mathematical Sciences, Institute of Natural Sciences, and MOE-LSC,}
   \centerline{Shanghai Jiao Tong University,}
   \centerline{ 800 Dongchuan Road, 200240, Shanghai, China}
} 

%

\bigskip

 \centerline{(Communicated by the associate editor name)}

\begin{abstract}
Granular materials are heterogenous grains in contact, which are ubiquitous in many scientific and engineering applications such as chemical engineering, fluid mechanics, geomechanics, pharmaceutics, and so on. Granular materials pose a great challenge to predictability, due to the presence of critical phenomena and large fluctuation of local forces. In this paper, we consider the quasi-static simulation of the dense granular media, and investigate the performances of typical minimization algorithms such as conjugate gradient methods and quasi-Newton methods. Furthermore, we develop preconditioning techniques to enhance the performance. Those methods are validated with numerical experiments for typical physically interested scenarios such as the jamming transition, the scaling law behavior close to the jamming state, and shear deformation of over jammed states.\end{abstract}

\section{Introduction}

Granular materials are conglomerates of discrete, macroscopic, solid particles, such as sand, soils, pills, etc. They are ubiquitous in many industrial and natural processes.
Analytical study is usually difficult due to the complicated nonlinear heterogeneous multi-body interactions in the dense granular system. Numerical methods have remarkable significance in the study of granular materials since most perturbative analytical methods in condense matter physics requires either the low density in the ideal gas limit or a prefect lattice for crystals. On the meanwhile, it is difficult for experimental methods to deal with large number of particles,  measure physical quantities in 3D and investigate phase transitions such as jamming phenomena quantitatively \cite{Kou:2017, Zhang:2014}. Numerical simulations allow access to all detailed information of the particle system, and usually it is straightforward to calculate all the relavant quantities (e.g. fabric, stress, energy) of the system.

However, it remains challenging to develop efficient and robust numerical methods for granular system, especially for large particle systems. Molecular dynamics and energy minimization are two main classes of simulation methods. Discrete element method (soft-particle method) is one of the typical molecular dynamics methods in granular applications \cite{Luding:2008}, where a set of Langevin equations of motion are formulated, with relaxation and random fluctuation terms to quantify the physical relaxation time scale and thermal effect, respectively. Related methods include event-driven method, non-smooth contact dynamics, etc, see references in \cite{Krijgsman:2016}. On the other hand, energy minimization methods can be applied to the situation of quasi-static deformation, when the applied load changes slowly over time with respect to the inertial forces. One thus can significantly speed up the simulation if time scales are not needed to be fully resolved. Also, the high accuracy of the energy minimization methods provides advantage for the quantitative study of the jamming transition and the scaling law behavior close to the jamming point. Although molecular dynamics method have been extensively used in the simulation of granular materials, the application of energy minimization methods is relatively recent. For example, Luding et. al. have applied trust region methods to study the relaxation and shear of granular system \cite{Krijgsman:2016}. In this paper, we will study the performance of several typical minimization methods for the granular simulation, such as conjugate gradient methods and quasi-Newton type methods. 

Preconditioning is the main bottleneck for the development of efficient and robust numerical algorithms for large scale molecular simulation \cite{Alrachid:2018, Packwood:2016}, boh in the case of molecular dynamics and energy minimization. General purpose preconditioners are not specifically targeted at large-scale atomistic systems, and are not particularly effective. In this paper, we will also stress on the development of preconditioning techniques for the efficient energy minimization of granular systems.

\begin{Remark}
The molecular details of molecular dynamics methods such as soft-particle methods and energy minimization methods are usually not the same. However, it is important to note that the equilibrium physical quantities such as fabric, stress and energy remain similar \cite{Krijgsman:2016}. Also, there is no temperature in the granular model simulated by energy minimization methods, namely, we are only interested in  athermal granular materials .
\end{Remark}

\subsection{Outline}
The paper is organized as follows. In Section \ref{sec:setup}, we present the physical model for the granular system, as well as the critical jamming transition which is important for the numerical investigation of dense granular system. We present the optimization and preconditioning techniques used for granular simulation in Section \ref{sec:numerics}. The numerical methods are validated and benchmarked in the numerical experiments in Section \ref{sec:numerical experiment}. We conclude in Section \ref{sec:conclusion} and point out some future improvements.

\section{Physical Setup}
\label{sec:setup}

\def\tot{\mathrm{tot}}

In this section, we introduce the basic physical models. The potential energy of the system is the sum of interaction energy of all particles. It is important to note that particles can interact unless they overlap. In other words, there is only repulsive force therefore the system is qualitatively different from the repulsive-attractive forces such as Lennard-Jones \cite{Edward:1924}. For simplicity, we only consider the contact models without tangential forces, and also only the two dimensional case.


\subsection{Granular configuration}
An admissible configuration $\C$ of particles is given by a list of center $\vec{x}_i$ and radius $r_i$ of particles. The volume (or packing) fraction $\phi$ is a key parameter in the simulation of  granular materials,
$$
\phi=\frac{1}{V}\sum_{i}\pi r_i^2
$$
where $V$ is the volume of the containing domain (container).


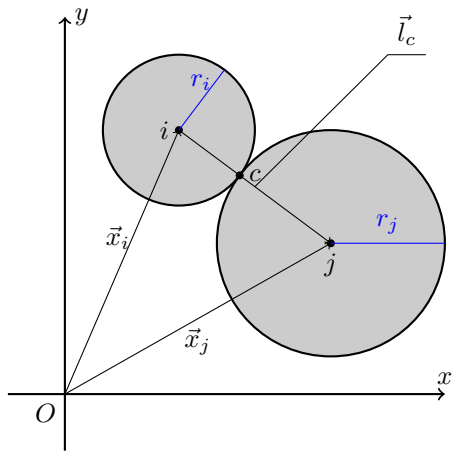
\begin{figure}[!htbp]
\begin{center}
\begin{tikzpicture}
\draw [black,thick,fill=black!20] (1.5,3.5) circle (1cm) node[left] {$i$};
\fill [black] (1.5,3.5) circle (1.5pt);
\draw [blue,-] (1.5,3.5) -- (2.1,4.3) node[above] at (1.79,3.91) {$r_{i}$};
\draw [black,thick,fill=black!20] (3.5,2) circle (1.5cm) node[below] {$j$};
\fill [black] (3.5,2) circle (1.5pt);
\fill [black] (2.3,2.9) circle (1.5pt) node[right] {$c$};
\draw [blue,-] (3.5,2) -- (5,2) node[above] at (4.25,2) {$r_j$};
\draw [thick, ->] (-0.75,0) -- (5,0) node[above] {$x$};
\draw [thick, ->] (0,-0.75) -- (0,5) node[right] {$y$};
\draw [black,->] (0,0) -- (1.5,3.5) node[above] at (0.7,1.8) {$\vec{x}_{i}$};
\draw [black,->] (0,0) -- (3.5,2) node[below] at (1.75,1) {$\vec{x}_{j}$};
\draw [black,->] (1.5,3.5) -- (3.5,2);
\path[draw] (2.5,2.75) -- (4.25,4.5) -- (4.75,4.5);
\node [above] at (4.5,4.5) {$\vec{l}_c$};
\node [below] at (-0.25,0) {$O$} ;
\end{tikzpicture}
\end{center}
\caption{Two particles in contact}
\label{fig:contact}
\end{figure}

The geometric structure of the granular materials can be characterized by the following fabric tensor,
\begin{equation}
F=\frac{1}{V}\sum_{c \in C}\pi\left(r_i^2 + r_j^2\right)\vec{n}_c\otimes \vec{n}_c
\end{equation}

\subsection{Contact Model}

Here we will introduce a contact potential of the following form,
\begin{equation} \label{eq: pair potential}
E_{ij} = \left\{
    \begin{aligned}
           & k_{ij}\frac{(r_i+r_j - \left|\vec{x}_i-\vec{x}_j\right|)^{\alpha}}{\alpha} &  &\text{for} \quad \left|\vec{x}_i-\vec{x}_j\right|< r_i+r_j, \\
           & 0 &  &\text{for} \quad \left|\vec{x}_i-\vec{x}_j\right|\ge r_i+r_j, ,
    \end{aligned}
\right.   i\ne j
\end{equation}
where $k_{ij}$ denotes the particle stiffness, $r_i$ and $r_j$ are the radii of particle $i$ and $j$, respectively, $\vec{x}_i$ and $\vec{x}_j$ are the positions of the their centers, respectively.

The total potential energy of the system is,
\begin{equation}\label{eq:Etot}
     E_{\tot} = \sum_{i<j}E_{ij}.
\end{equation}
See Figure \ref{fig:contact} for an illustration.

Since the particles interact with each other only if they contact, we can rewrite the potential energy for a single contact $c$ as,
\begin{equation}
E_c = k_c\frac{(\delta_c\vee 0)^{\alpha}}{\alpha},
\end{equation}
where $\delta_c$ is the overlap associated with the contact $c$, and by equation \eqref{eq: pair potential},
\begin{equation}
\delta_c = r_i + r_j - \left|\vec{x}_i-\vec{x}_j\right|.
\end{equation}
In this formula, particle $i$ and $j$ are two particles associated with contact point $c$ (see Figure \ref{fig:contact}).  Therefore, if $\delta_c\geq 0$, the first derivatives, or forces of $E_c$ are,
\begin{equation} \label{eq: first derivative}
  \begin{aligned}
    & \frac{\partial E_c}{\partial \vec{x}_i} = -\vec{f}_c = -k\delta_c^{\alpha-1}\vec{n}_c \\
    & \frac{\partial E_c}{\partial \vec{x}_j} = \vec{f}_c = k\delta_c^{\alpha-1}\vec{n}_c,
  \end{aligned}
\end{equation}
where
\begin{equation}
\vec{n}_c = \frac{\vec{l}_c}{l_{c}} ,\quad \vec{l}_c = \vec{x}_i-\vec{x}_j, \quad l_c = \left| \vec{l}_c \right|,
\end{equation}
and $\vec{f}_c$ is the contact force acting on particle $i$ by particle $j$.
Furthermore, the second derivatives are,
\begin{equation} \label{eq: second derivative}
  \begin{aligned}
    & \frac{\partial^2 E_c}{\partial\vec{x}_i \partial\vec{x}_i} = k\delta_c^{\alpha-2} \left(\left(\alpha-1+\frac{\delta_c}{l_c}\right)\vec{n}_c\otimes \vec{n}_c - I \frac{\delta_c}{l_c} \right) \\
    & \frac{\partial^2 E_c}{\partial\vec{x}_j \partial\vec{x}_j} = \frac{\partial^2 E_c}{\partial\vec{x}_i \partial\vec{x}_i}\\
    & \frac{\partial^2 E_c}{\partial\vec{x}_i \partial\vec{x}_j} = -\frac{\partial^2 E_c}{\partial\vec{x}_i \partial\vec{x}_i},
  \end{aligned}
\end{equation}
where $\otimes$ is the tensor product of two vectors and $I$ is the $2\times 2$ identity matrix.

\begin{remark}
The exponent $\alpha\geq 2$ identifies the nonlinearity of the pair potentials. When $\alpha=2$, the interaction between particles follows the \emph{Hooke's law}, which models a repulsive harmonic spring. The force between two overlapped particles linearly depends on the overlap. Hooke's model is one of the simplest contact model, which is still of great importance. From equation \eqref{eq: first derivative} and \eqref{eq: second derivative}, it is clear that the first derivatives of the Hooke potential are continuous at $\delta_c = 0$, while the second derivatives are discontinuous there.
When $\alpha > 2$, we obtain the so-called \emph{Modlin-Hertzian model} or \emph{Hertzian model}, which is based on the Hertz contact theory \cite{Johnson:1987}. The force between two overlapped particles has a power law nonlinearity.
Both the first and second derivatives of the Hertzian potential are continuous if $\alpha >2$. We take $\alpha = 5/2$ in the numerical examples of this paper.
\end{remark}

\subsection{Macroscopic Variables}

The average stress tensor can be calculated by the following formula,
\begin{equation} \label{eq: def of stress tensor}
\sigma=\frac{1}{V}\sum_{c \in C}\vec{l}_c\otimes \vec{f}_c,
\end{equation}
where $V$ is the volume of the containing domain and $C$ is the collection of all the contact points in the domain. According to equation \eqref{eq: def of stress tensor}, the pressure of the system can be defined as,
\begin{equation}
p=\frac{\sum_{\alpha=1}^d\sigma_{\alpha\alpha}}{d}
\end{equation}
where $d$ is the dimension of the domain. The bulk modulus can then be defined as,
\begin{equation}
B=\phi\frac{dp}{d\phi}.
\end{equation}

For a simple shear \cite{Ogden:1997} such that the deformation matrix is given by
\begin{displaymath}
D=\left(
    \begin{array}{cc}
      1 & \gamma \\
      0 & 1 \\
    \end{array}
  \right),
\end{displaymath}
the shear stress is $-\sigma_{12}$, and the shear modulus can be calculated by: 
\begin{equation}\label{eq: cal shear modulus}
G = -\frac{d\sigma_{12}}{d\gamma}
\end{equation}

In addition, the average contact number or coordination number, denoted by $\bar{z}$, is a key geometric parameter in granular simulation. For a granular system consisting of $N$ particles,  $\bar{z}$ can be simply defined by,
$$
\bar{z}=\frac{|C|}{N}
$$
where $|C|$ is the number of all contact points.

We note that there exist particles with zero contacts, namely, the so called "rattlers". In addition, in our 2D frictionless circular disk system, some particle are called "dynamic rattlers" \cite{Fatih:2010} if their numbers of contacts are less than 3, which can lead to mechanical instability. Thus, these rattlers are excluded from the calculation of the coordination number. We denote the number of particles with at least 3 contacts by $N_3$, and the set of contact points for particles with at least 3 contacts by $C_3$. The modified definition of coordination number is,
\begin{equation}
\bar{z}=\frac{|C_3|}{N_3}
\label{eq:z3}
\end{equation}
\subsection{Jamming formation}


The jamming transition is an important physical process in granular system \cite{Majmudar:2005, O'Hern:2003, Zhang:2005}, which refers to the critical slowing down of the system due to "overcrowded" particles. Close to the transition point, the geometric constraints prevent the system to explore the phase space, and the system changes its behavior from gas or liquid like to solid like. During the transition, the system has to adjust its configurations more and more in a collective manner instead of via local movements, in order to achieve energy equilibrium. This corresponds to the essential difficulty for the numerical simulation which will be discussed later.


In this paper, we adopt the approach by O'Hern et al. \cite{O'Hern:2003} to generate a jamming configuration: An initial configuration is sampled with sufficiently low packing fraction $\phii$ such that there is no overlapping particles after relaxation (energy minimization). The total energy and pressure for the relaxed system with volume fraction $\phii$ is $0$. Starting from $\phii$, we repeatedly increase the volume fraction $\phi$ with a small increment $\dphi$ by enlarging the size of particles accordingly, and then carry out energy relaxation. At some critical fraction $\phic$, there will be no free space for particles to explore and unavoidably, they will come into contact. As the system is further compressed, some particles overlap, and the total energy and pressure become nonzero and starts to grow. The system will also posses nonzero bulk modulus since the pressure increases upon compression. $\phic$ is called the critical volume fraction or jamming point. In  2D bi-disperse system with number ratio 1:1 and size ratio 1.4:1, $\phic$ is approximately 0.842 \cite{Krijgsman:2016}.


\def\app{\mathrm{app}}
\def\pre{\mathrm{pre}}
We can adopt a bisection algorithm to increase the accuracy of $\phic$. Let $\phic^\app$ be an approximate critical fraction associated with a configuration with nonzero energy and pressure, the previous unjammed configuration with zero energy and pressure has the volume fraction $\phic^{\pre}:=\phic^\app - \dphi$. We can then compress the unjammed configuration at $\phic^{\pre}$ by increasing its volume fraction to $\phim = \phic^{\pre} + \dphi/2$, and see if the jamming transition already happens at $\phim$. This procedure can be repeated until for example, $\dphi< \varepsilon$, where $\varepsilon$ represents the desired accuracy. We note that this procedure can achieve very high accuracy for the determination of the jamming point, which is not feasible for experimental methods.

The overall procedure is summarized in the following algorithm for the generation of the jamming configuration. Denote by $\min E(\phi, \Cr)$ the energy minimization in the configuration space $\C$ with the packing fraction $\phi$ and the initial configuration $\Cr$, which can be solved by certain optimization algorithm in Section \ref{sec:numerics}. The minimizer is denoted by
$\C_{\mathrm eq} = \argmin E(\phi, \Cr)$.
Here we use a square shaped container $\Omega$ and the periodic boundary condition for the granular system.
\def\pref{p_{\mathrm{ref}}}
\def\Cm{\C_m}
\def\Cref{\C_{\mathrm{ref}}}
\def\Eeq{E_{\mathrm{eq}}}
\def\peq{p_{\mathrm{eq}}}

\renewcommand{\algorithmicrequire}{ \textbf{Input:}} 
\renewcommand{\algorithmicensure}{ \textbf{Output:}} 

\begin{algorithm}[H]
\caption{Generation of the Jamming Configuration.}
\label{alg:jamming}
\begin{algorithmic}[1]
\REQUIRE ~~\\
particle number $N$;\\
initial packing fraction $\phii$;\\
increment $\dphi$; \\
prescribed accuracy $\varepsilon$.
\ENSURE ~~\\
critical volume fraction $\phic$;\\
jamming configuration $\Cj$ with $\phic$.
\STATE Generate an initial configuration $\Ci$ with $N$ particles and volume fraction $\phii$;
\STATE let $\Cref = \argmin (\phii, \Ci)$, with total energy $\Eref=0$ and pressure $\pref=0$.
\label{alg:jam:step1}
\STATE // Increment Step
\WHILE{$\Eref=0$ and $\pref=0$}
\STATE Let $\phi = \phir + \dphi$;\\
\STATE Fix the position of the particles, and enlarge their radii uniformly to generate a new configuration $\Cr$ with volume fraction $\phi$;\\
\STATE Let $\Ceq(\phi) = \argmin E(\phi, \Cr)$,  compute $E(\Ceq)$ and $p(\Ceq)$.
\IF{$E(\Ceq)>0$ \& $p(\Ceq)>0$}
\STATE  $\phic^\app = \phi$, $\Cj = \Ceq$, break;
\ELSE
\STATE  $\phir = \phi$, $\Cr = \Ceq$;
\ENDIF
\ENDWHILE
\STATE  // Bisection Step
\WHILE{$\phic^\app - \phir > \epsilon$ }
\STATE  $\phim= (\phic + \phir)/2$;
\STATE  Generate an intermediate configuration $\Cm$ with volume fraction $\phim$, starting from $\Cr$.
\STATE  Let $\Ceq = \argmin (\phim, \Cm)$, and calculate $E(\Ceq)$ and $p(\Ceq)$.
\IF{$E(\Ceq)>0$ \& $p(\Ceq)>0$}
\STATE   $\phic^\app = \phim$;
\ELSE
\STATE $\phir = \phim$, $\Cr = \Ceq$;
\ENDIF
\ENDWHILE

\STATE Denote $\phic = \phic^\app$, $\Cj = \Ceq$.
\RETURN $\phic$, $\Cj$;
\end{algorithmic}
\end{algorithm}

\section{Numerical Method}
\label{sec:numerics}

\newcommand{\br}[1]{{(#1)}}

The minimization of the total potential energy \eqref{eq:Etot} relies on the state-of-the-art optimization and precondition techniques which will be presented in this section. The performance of those methods is contingent upon the particular applications. In particular, the construction of a good preconditioner, can be regarded as "an art rather than a science" \cite{Wathen:2015}.

\subsection{Energy minimization}

In this paper, we use the limited-memory Broyden–Fletcher–Goldfarb–Shanno (L-BFGS) method \cite{Nocedal:1980} and Fletcher-Reeves conjugate gradient (FR-CG) method \cite{Nocedal:2006} for the energy minimization of granular system. We denote the total energy of a system as $f(x)$ where $x$ is the positions of all particles. If $x^\br{k}$ is the iterate at step $k$, we denote the energy at step $k$ by $f^\br{k} = f(x^\br{k})$, and the gradient at step $k$ by $g^\br{k}=\nabla f^\br{k}$.

\subsubsection{L-BFGS method}
L-BFGS is one type of the quasi-Newton methods, namely, it utilizes an approximation of the inverse Hessian matrix to generate the search direction. Instead of the full  dense $n\times n$ approximations to the inverse Hessian ($n$ is the number of variables, which is $Nd$ with $N$ number of particles and $d$ the dimension in our problem), L-BFGS uses a low rank approximation with only a few vectors of length $n$ to represent the approximation implicitly.

We denote the approximate inverse Hessian at step $k$ by $B^{(k)}$. Assuming we have stored the last $m$ updates of the form $s^\br{k}=x^\br{k+1}-x^\br{k}$ and $y^\br{k}=g^\br{k+1}-g^\br{k}$, we define $\displaystyle\rho^\br{k}=\frac{1}{y^{(k),T}s^\br{k}}$.
The L-BFGS search direction $p^{\br{k}} (= B^{(k)}\nabla f^\br{k})$ can be obtained using the following algorithm.

\begin{algorithm}[H]
\caption{L-BFGS two loop recursion.}
\label{alg:L-BFGS}
\begin{algorithmic}[1]

\STATE $q=g^{(k)}$\;

\FOR{ $i=k-1,k-2,...,k-m$}
\STATE $\alpha^\br{i}=\rho^\br{i} s^{(i),T} q$\;
\STATE $q=q-\alpha^\br{i} y^\br{i};$
\ENDFOR
\STATE $\boxed{r=B^{(k)}_0q}$\;

\FOR{$i=k-m,k-m+1,...,k-1$}

\STATE $\beta=\rho^\br{i}y^{(i),T}r$\;
\STATE $r=r+s^\br{i}(\alpha^\br{i}-\beta)$\;
\ENDFOR

\RETURN $p^\br{k} = r (= B^\br{k}\nabla f^\br{k})$\;
\end{algorithmic}
\end{algorithm}

The matrix $B^{(k)}_0$ in the boxed step in the algorithm is an rough approximation to the inverse of the exact Hessian, and an effective choice is:
$$
B^{(k)}_0=\frac{s^{(k-1),T}y^\br{k-1}}{y^{(k-1),T}y^\br{k-1}}I
$$
This choice ensures that the search direction is well scaled and therefore the unit step length is accepted in most iterations. Besides, the diagonal matrix makes it much simpler to compute the multiplication $r=B^{(k)}_0q$. 


\subsubsection{FR-CG method}

We use the Fletcher-Reeves variant \cite{Fletcher:1964} of the nonlinear conjugate gradient (CG) method \cite{Pytlak:2008}. As an extension of the linear conjugate gradient method \cite{Hestenes:1952}, the search direction $p^{(k)}$ at $k$th step of FR-CG method is defined by,

\def\FR{\mathrm{FR}}
\begin{algorithm}[H]
\caption{FR-CG.}
\label{alg:FRCG}
\begin{algorithmic}[1]

\STATE $\displaystyle\beta_{\FR}^{\br{k}} = \frac{g^{\br{k}, T}{g^{\br{k}}}}{g^{\br{k-1}, T}{g^{\br{k-1}}}}$;
\RETURN $p^{\br{k}} = -g^{(k)} + \beta_{\FR}^{\br{k}} p^\br{k-1}$.
\end{algorithmic}
\end{algorithm}

\subsubsection{Line Search}

Once we calculate the new search direction using either L-BFGS in Algorithm \ref{alg:L-BFGS} or FR-CG in Algorithm \ref{alg:FRCG}, the next iteration is given by
\begin{equation}
x^\br{k+1}=x^\br{k}+\alpha^\br{k}p^\br{k},
\end{equation}
where $\alpha^\br{k}$ is chosen by a line search method which satisfies Wolfe conditions or strong Wolfe conditions \cite{Nocedal:2006}, such that the update is stable.

\subsection{Preconditioning}
Preconditioning is important for the efficiency of the minimization algorithms, especially for large scale problems. We have to choose the preconidtioner matrix $P^\br{k}$ such that it is similar to the exact Hessian, and at the same time easy to invert.

\subsubsection{Preconditioned minimization algorithm}
For L-BFGS algorithm \ref{alg:L-BFGS}, we replace the boxed step in the algorithm with:
\begin{equation}\label{eq: preconditioned step}
\boxed{r = P^{(k),-1}q}
\end{equation}
in order to obtain a preconditioned L-BFGS algorithm.

For FR-CG algorithm \ref{alg:FRCG}, we have the following preconditioned version for $p^{\br{k}}$.
\def\FR{\mathrm{FR}}
\begin{algorithm}[H]
\caption{Preconditioned FR-CG method.}
\label{alg:PFRCG}
\begin{algorithmic}[1]
\STATE $y^{\br{k}} = P^{\br{k},-1} g^{\br{k}}$;
\STATE $\displaystyle\beta_{*\FR}^{\br{k}} = \frac{y^{\br{k}, T}{g^{\br{k}}}}{y^{\br{k-1}, T}{g^{\br{k-1}}}}$;
\RETURN $p^{\br{k}} = -y^{(k)} + \beta_{*\FR}^{\br{k}} p^\br{k-1}$.
\end{algorithmic}
\end{algorithm}

The equation \eqref{eq: preconditioned step} or Step 1 in Algorithm \ref{alg:PFRCG} is equivalent to solve a linear system:
\begin{equation}\label{eq:solvpsys}
P^\br{k}r=q.
\end{equation}

An alternative point of view for the preconditioning is to make a change of coordinates: $\wx= P^{(k),1/2}x$. Considering the function $F(\wx)=f(P^{(k),-1/2}\wx)$, we have
\begin{equation}
\nabla F(\wx)=P^{(k),-1/2}\nabla f(P^{(k),-1/2}\wx).
\end{equation}

Applying the L-BFGS algorithm \ref{alg:L-BFGS} or FR-CG algorithm \ref{alg:FRCG} to optimize the transformed function $F(\tilde{x})$, we can obtain the preconditioned version of the corresponding algorithm.

\subsubsection{Construction of preconditioners for large scale molecular system}

\def\cut{\mathrm cut}
\def\stab{\mathrm stab}

Preconditioning is well established in numerical linear algebra and numerical PDE problems \cite{Wathen:2015}. However, in many applications general purpose preconditioners do not work particularly well, and specifically designed preconditioned are preferable. For large scale atomic/molecular simulations, Packwood et. al. proposed the so-called "universal preconditioner" $P$ \cite{Packwood:2016}. They have successfully applied this preconditioner to atomic simulation and electronic structure calculation of crystalline system of the size from hundreds to $10^4$ atoms, with a speedup of order 1 or 2 \cite{Mones:2018}. In this paper, we will test the performance of those universal preconditioners for granular systems. 
 
 The universal preconditioner $P$ is defined via the quadratic form 
\begin{equation}\label{eq: precond matrix entry}
\begin{aligned}
	u^T P u &= \mu \sum_{0<|r_{ij}|<r_{\cut}} c_{ij} |u_i - u_j|^2, \\
        c_{ij} &= \exp(-A (r_{ij}/r_{nn} -1)). 
        \end{aligned}
\end{equation}
Here $r_{\cut}$ is the cutoff radius, $\mu$ is the energy scale, 
$A$ is a parameter which should be large enough to ensure that the nearest neighbor interactions dominant. The numerical experiments in \cite{Packwood:2016} and in our numerical results indicate that as long as $A$ is of order 1, it does not significantly affect the performance of the preconditioner. In particular, if we take $A=0$, we choose the preconditioner matrix $P$ as the stabilized adjacency matrix of the particles \cite{Packwood:2016}, namely,
\begin{equation}\label{eq: precondition matrix}
\begin{aligned}
P_{ij} &= \left\{
    \begin{aligned}
            -\mu, & &  d_{ij}< r_{\cut} \\
            0, & &  d_{ij}\ge r_{\cut}
    \end{aligned},
   \right.\\
P_{ii} &= -\sum_{i \ne j}P_{ij} + \mu C_{\stab},
\end{aligned}
\end{equation}
where $d_{ij}$ is the distance between particle $i$ and $j$, $C_{\stab}$ is a stabilization term to make sure the matrix is strictly positive definite, we choose $C_{\stab}=0.1$ in our application.  

In our model problem, the bi-disperse granular system is composed of two types of particles, whose radii are denoted by $r_{\min}$ and $r_{\max}$. We take the cutoff radius $r_{\cut} > 2r_{\max}$ so that all the possibly overlapping particles are covered. For example, we can take  $r_{\cut} = 2.2r_{\max}$. The preconditioner matrix will be recalculated when the maximum displacement of particles is beyond $r_{cut}/4$, therefore it will not be recalculated frequently. The energy scale $\mu$ is chosen to make sure that the precondition matrix is of the same order as the real hessian, so we may choose the unit step-length when the L-BFGS method is applied. One way to obtain $\mu$ is,
\begin{equation}
\nu^{T}\left(\nabla E(x_{0}+\nu)-\nabla E(x_{0})\right)=\mu \nu^{T}P_{\mu=1}\nu,
\end{equation}
where $P_{\mu=1}$ is the matrix when $\mu=1$, and $\nu$ is a test perturbation with the following form
\begin{equation}
\nu(x)=M\left(\text{sin}(x/L)\right),
\end{equation}
where $L$ is the lengths of the periodic cell and $M$ is a constant, here we choose $M=10^{-3}r_{cut}$. 

Since we need to solve linear system \eqref{eq:solvpsys} with respect to $P$, when the particle number is small, we use direct method and rearrange the index of grains by sparse reverse Cuthill-McKee ordering method \cite{Cuthill:1969}, the band width of $P$ can be reduced so that the equation \eqref{eq:solvpsys} can be easily solved. If the number of particles is large, iterative solvers such as preconditioned CG or AMG \cite{Brandt:1985} can be utilized. 


\section{Numerical Results}
\label{sec:numerical experiment}

In our numerical experiments, we use the bidisperse granular system which consists of two types of particles with number ratio 1:1 and  size ratio $r_{\max}/r_{\min} = 1.4$, so the system will not crystallize \cite{Speedy:1999}. The particles are contained in a unit square domain $\Omega = [0,1]^2$. We use the periodic boundary condition in order to reduce the boundary layer effect. Solid wall boundary conditions can also be implemented.

In the numerical experiments, we test the performance of preconditioned L-BFGS method and preconditioned FR-CG method, by running several benchmark simulations: the jamming formation, the scaling law behavior for the jammed configuration, and the shear deformation of the over-jammed states. In those different scenarios, we will show that the proposed preconditioned algorithm has better performance compared to their un-preconditioned counterparts, and preconditioned L-BFGS is currently the method of choice.

\subsection{Jamming formation}

We first run Algorithm \ref{alg:jamming} to generate the jamming configuration, starting from a relaxed configuration with volume fraction $\phi< \phic$, where $\phic$ is the critical volume fraction (or jamming fraction). We let the particles grow with small increment $\delta\phi$, then we use either the preconditioned or un-preconditioned version of Algorithm \ref{alg:L-BFGS} and Algorithms \ref{alg:FRCG}, \ref{alg:PFRCG} as the energy minimizer in Algorithm \ref{alg:jamming}. 

\begin{figure}[H]
  \centering
  \includegraphics[width=0.55\linewidth]{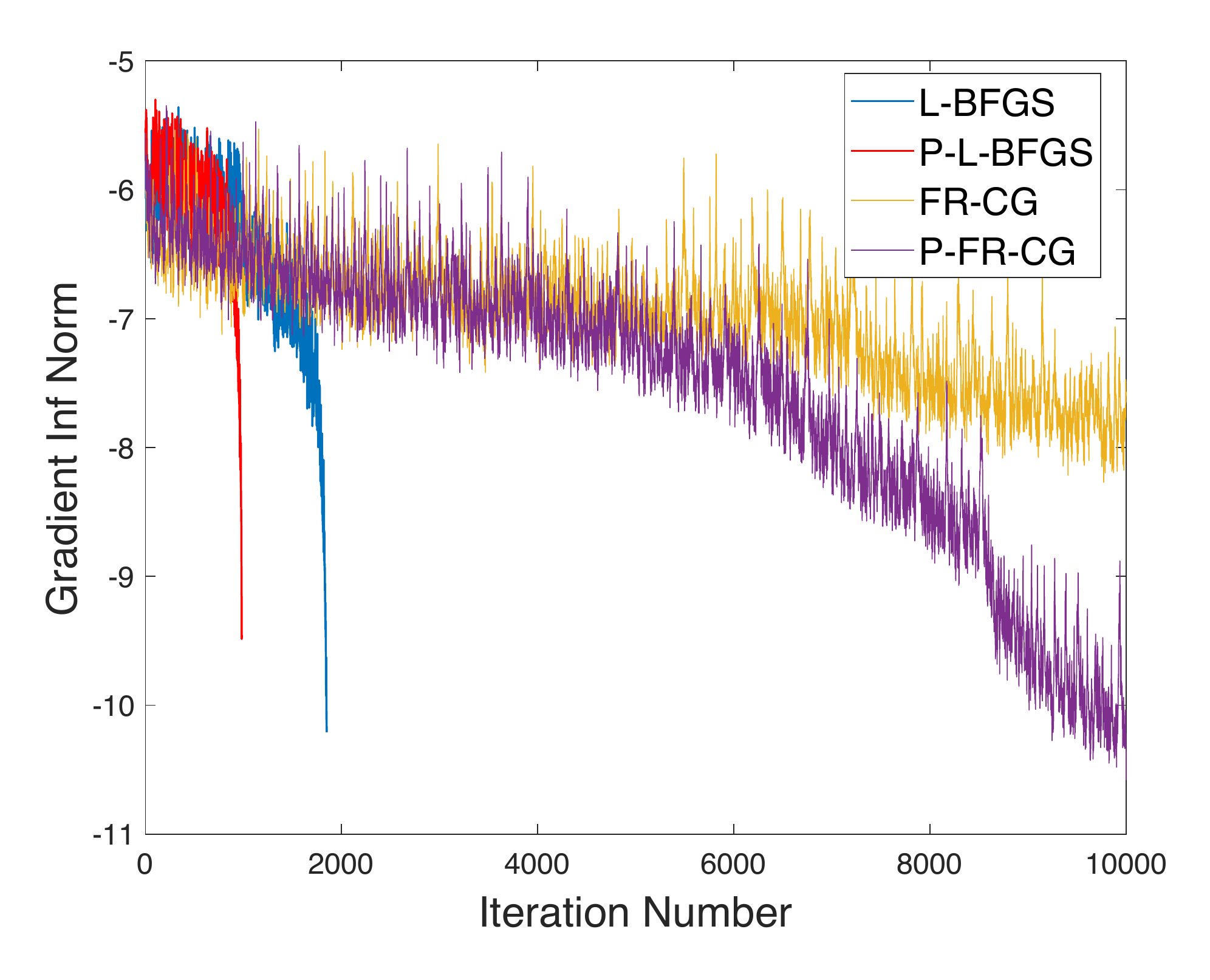}\\
  \caption{$\ell^\infty$ norm of gradient vs iteration number for one energy minimization step of a granular system with particle number 4096, $\phi=0.8389$, the radii of particular grow by $r \rightarrow (1+\delta)r, \delta=5\times 10^{-5}$.}\label{fig: gradient vs iteration all method}
\end{figure}

 In Figure \ref{fig: gradient vs iteration all method} , the residual norm of each iteration during a relaxation process is plotted. We apply four methods (L-BFGS, preconditioned L-BFGS, FR-CG, Preconditioned FR-CG) to minimize the energy of an unjamming configuration. A high accuracy tolerance is adopted here when an accurate jamming point is pursued. The figure shows that: L-BFGS is better than FR-CG, preconditioned L-BFGS method has the best convergence curve compared to other methods, and preconditioner provides a factor of two speed up for L-BFGS. 

We also notice that convergence curves for all methods are extremely rough and have long asymptotic regimes, that means the energy landscape is very complex and there are many local minimizers to explore.

\begin{figure}[H]
\centering
\subfigure[]{
\begin{minipage}[t]{0.33\linewidth}
\centering
\includegraphics[width=0.9\linewidth]{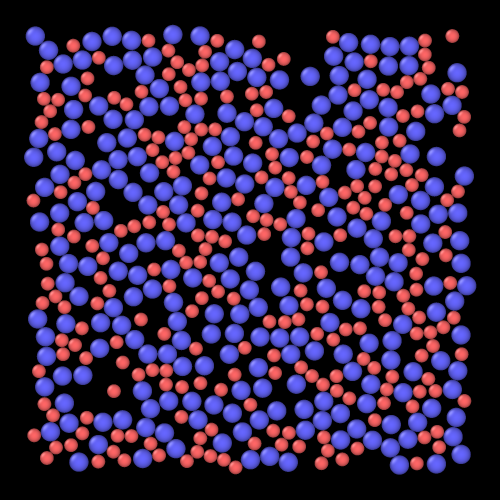}
\end{minipage}%
}%
\subfigure[]{
\begin{minipage}[t]{0.33\linewidth}
\centering
\includegraphics[width=0.9\linewidth]{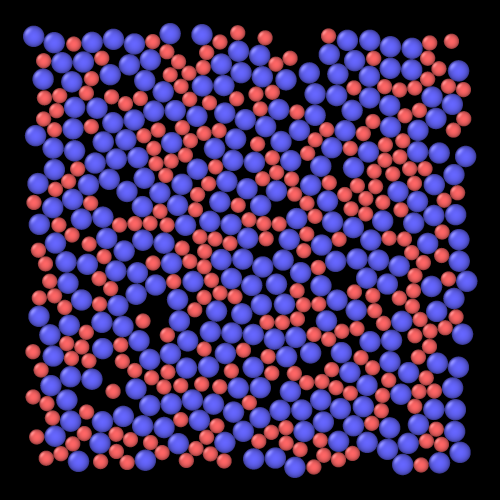}
\end{minipage}%
}%
\subfigure[]{
\begin{minipage}[t]{0.33\linewidth}
\centering
\includegraphics[width=0.9\linewidth]{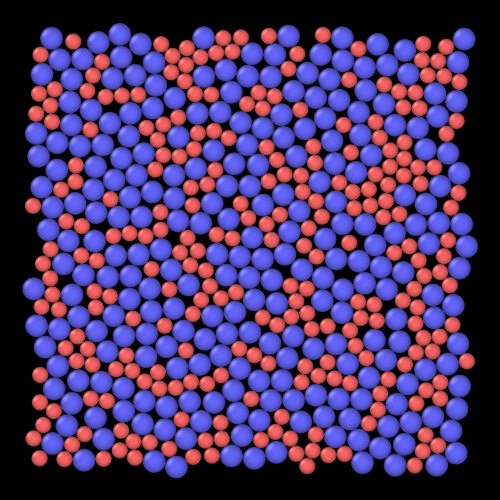}
\end{minipage}
}%
\centering
\caption{Schematic of the evolution from non-jamming to jamming state of bi-disperse granular system using Algorithm \ref{alg:jamming}, \textbf{a} Initial non-jamming configuration. \textbf{b} Intermediate configuration. \textbf{c} Jamming configuration, \emph{Visualization tool: OVITO \cite{Stukowski:2009}}.}\label{fig: jamming progress}
\end{figure}

Figure \ref{fig: jamming progress} shows the evolution of the granular system up to the jamming point. The volume fraction of the initial configuration is small, therefore particles can move around in order to achieve a zero energy configuration. When the volume fraction increases, the particles start to contact with each other, but initially it is still possible to achieve a zero energy configuration through energy relaxation (minimization). As the free space becomes less and less, at a critical volume fraction, the particles are forced to contact and overlap, and the potential energy of the system becomes non-zero.

\subsection{Scaling Law.}
Once the jamming configuration $\Cj$ is obtained, we can continue increasing the volume fraction in a quasi-static manner. It has been proposed by physicists that the granular system has the so-called scaling law behavior close to the jamming point \cite{O'Hern:2003}. Here, we numerically justify this scaling law using our energy minimization techniques, and more interestingly, we observe that the iteration number of the energy minimization methods also has similar scaling law behavior.

\begin{figure}[H]
  \centering
  \includegraphics[width=0.55\linewidth]{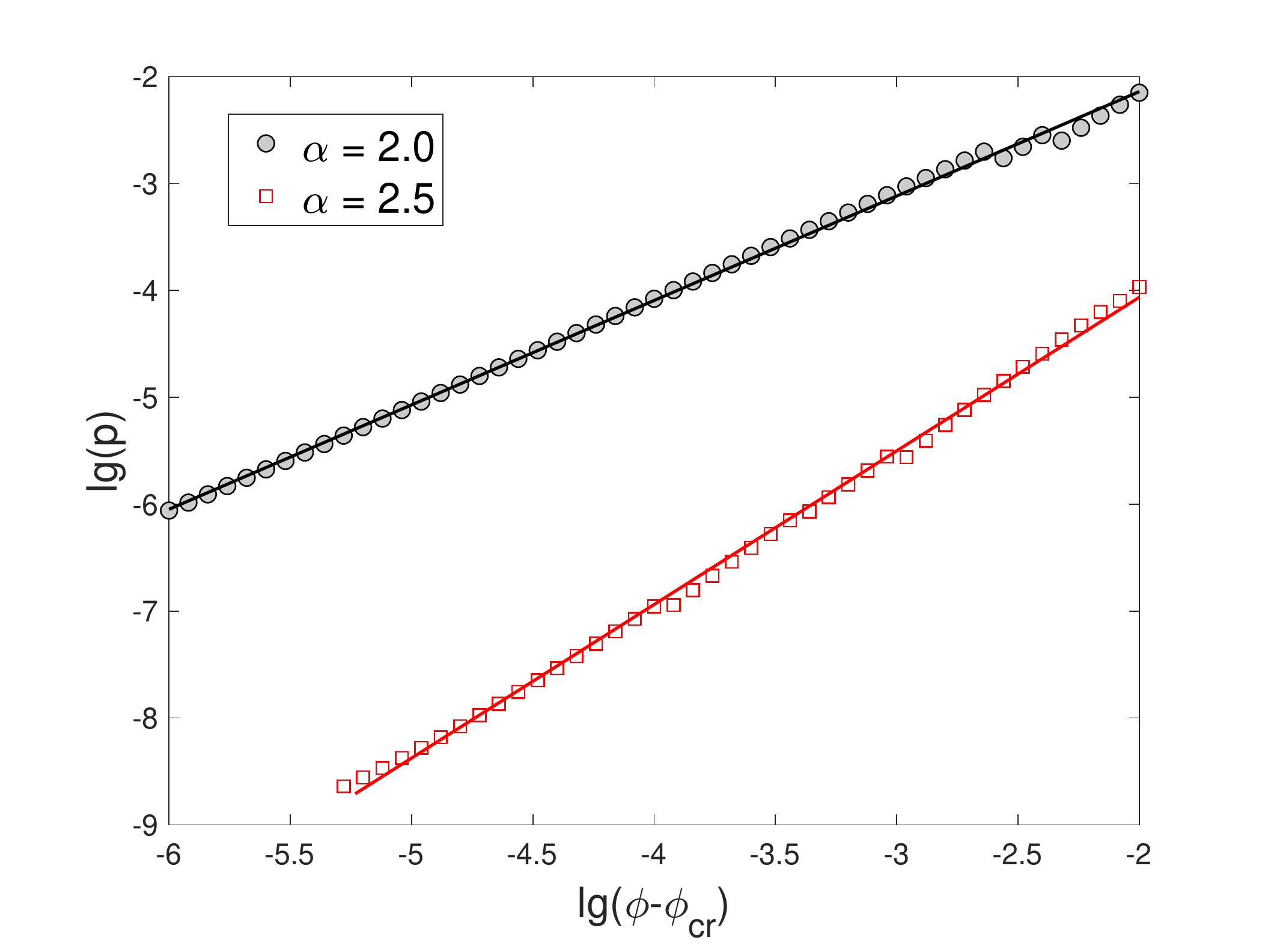}\\
  \caption{pressure $p$ vs $\phi-\phic$ for a 2D bi-disperse granular system with particle number 4096.}\label{fig:pvsphi}
\end{figure}

In the Figure \ref{fig:pvsphi}, we plot the log-log curve of pressure $p$ vs $\phi-\phic$. It is clear that $p$ follows a power law with respect to $\phi-\phic$ close to the jamming point, and the critical exponent depends on the nature of interactions. For harmonic $(\alpha=2)$ and Hertzian $(\alpha=2.5)$ potentials, the critical exponent are approximately $1.0$ and $1.5$, respectively. In addition, this scaling law between $p$ and $\phi-\phic$ is robust for different initial configurations (they may have different $\phic$).

\begin{figure}[H]
  \centering
  \includegraphics[width=0.55\linewidth]{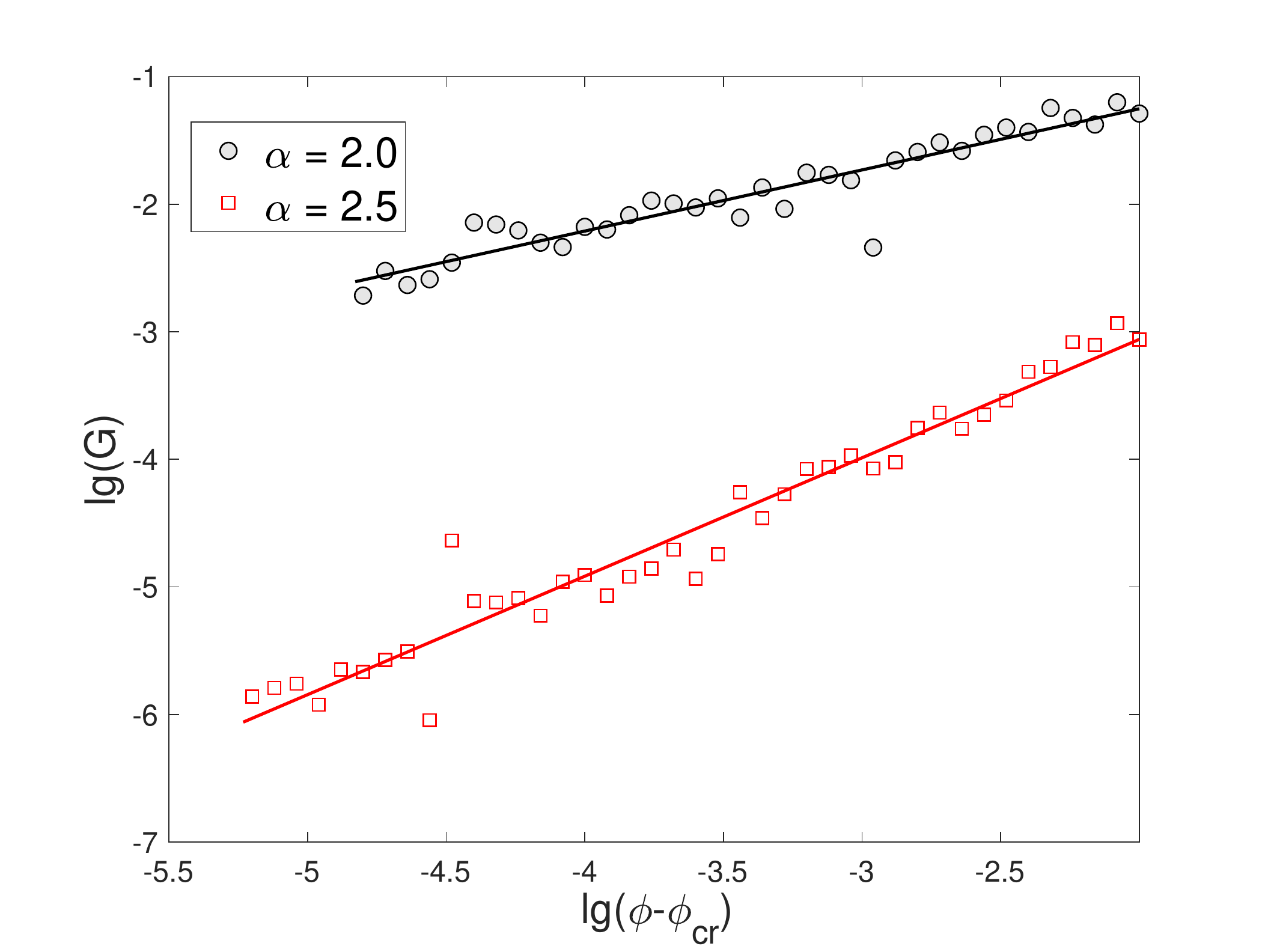}\\
  \caption{Shear modulus $G$ vs $\phi-\phic$ for a 2D bi-disperse granular system with particle number 4096.}\label{fig:GvsPhi}
\end{figure}

If we exert small simple shear strain $\gamma$ to these configurations, we can compute their shear modulus using equation \eqref{eq: cal shear modulus}. In Figure \ref{fig:GvsPhi}, we plot shear modulus for different volume fractions. It shows that the shear modulus $G$ of the granular systems also follows a similar power scaling law with respect to $\phi-\phic$ but with different exponents, which is approximately $0.5$ when $\alpha=2.0$, and $1.0$ when $\alpha=2.5$.


\begin{figure}[H]
  \centering
  \includegraphics[width=0.55\linewidth]{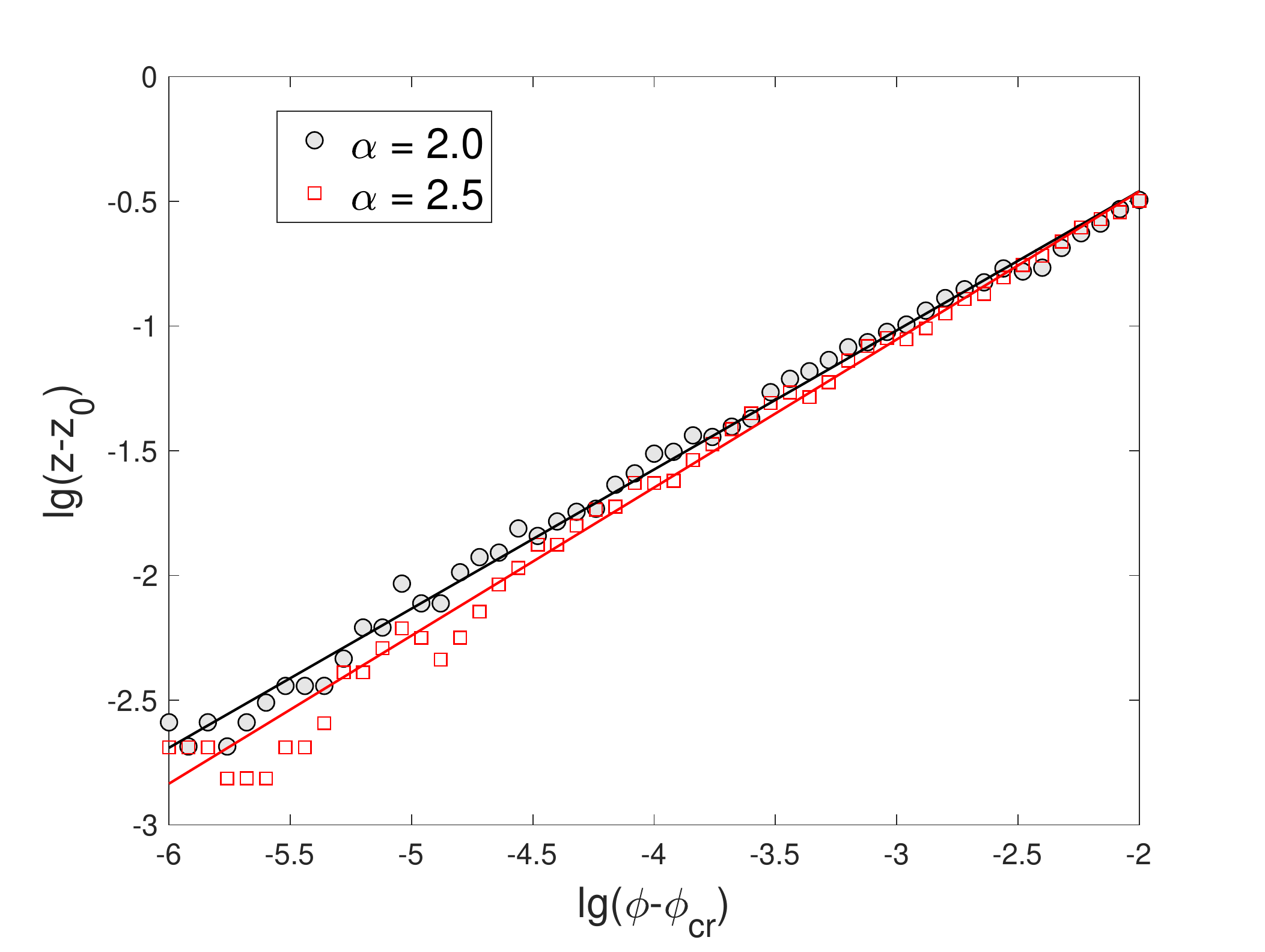}\\
  \caption{Average contact number $\bar{z}$ vs $\phi$ for a 2D bi-disperse granular system with particle number 4096. $z_0$ is the average contact number of the granular system with volume fraction $\phic$. In our 2D simulation, particles are frictionless, $z_0=4.0$.}\label{fig:zvsPhi}
\end{figure}

The average contact number $\bar{z}$ defined in \eqref{eq:z3} also exhibits a scaling law in regards to volume fraction difference $\phi-\phic$. In Figure \ref{fig:zvsPhi}, the lines in the log-log plots have similar slopes close to $0.5$ for both $\alpha=2.0$ and $\alpha=2.5$. The critical average contact number $z_0=4$ is the average contact number when a system reaches exactly the jamming state. The average contact number of a system takes a leap from zero to 4 when jamming transition occurs.

\begin{figure}[H]
  \centering
  \includegraphics[width=0.6\linewidth]{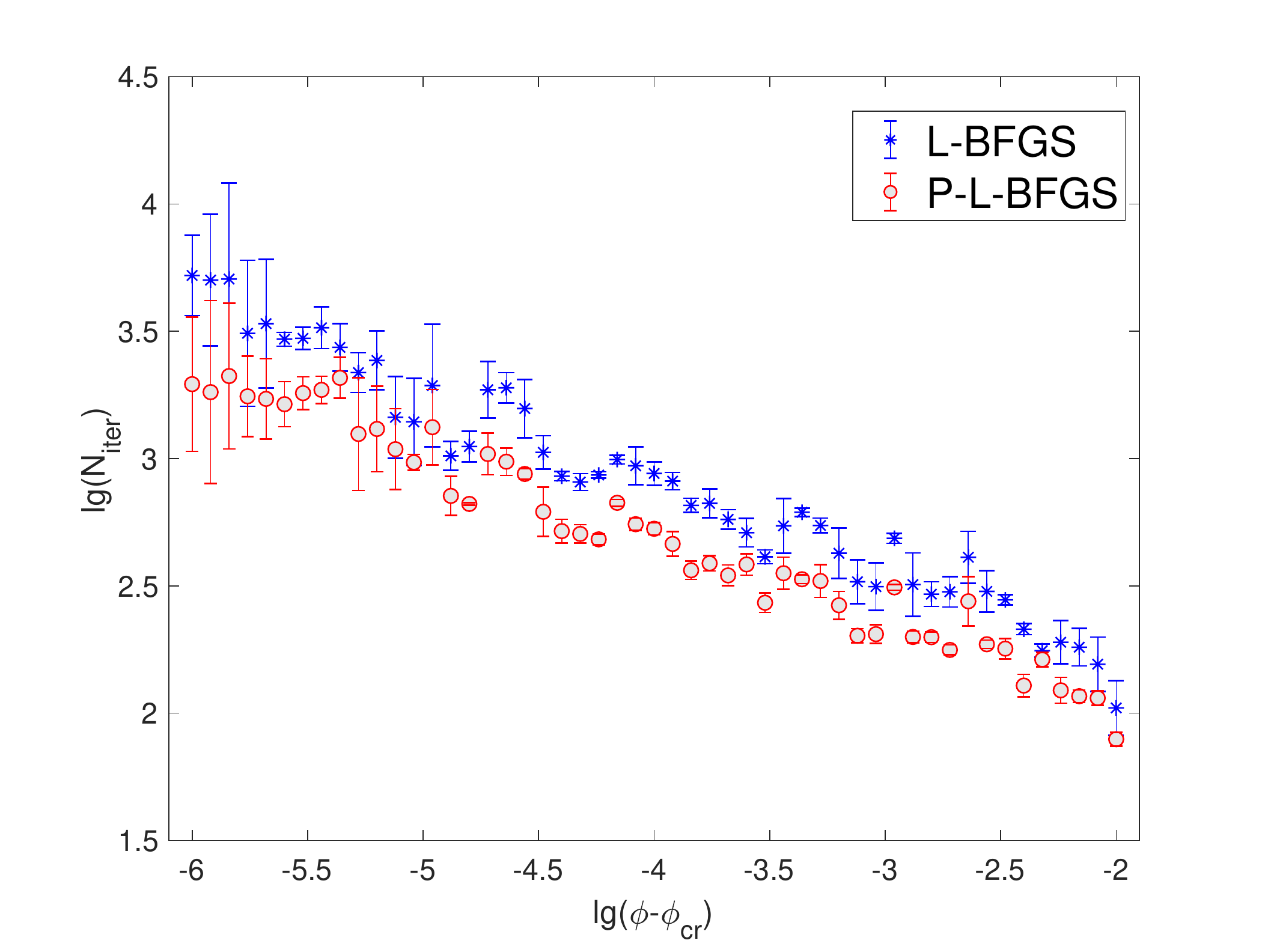}\\
  \caption{Iteration numbers (average and variance) with respect to $\phi-\phic$. Red circles and blue stars are average iteration numbers of 10 quasi-static simulations using preconditioned L-BFGS method and un-preconditioned L-BFGS method, respectively. In addition, we plot the error bars for those simulations.}
  \label{fig:itervsphi}
\end{figure}

In Figure \ref{fig:itervsphi}, we plot the iteration number vs. volume fraction difference $\phi - \phic$ for the preconditioned L-BFGS and L-BFGS methods. Starting from certain jamming configuration, we compress the granular system by increasing the volume fraction $\phi$. Then for each relaxed over-jammed configuration, we exert small shear strains from $10^{-6}$ to $10^{-5}$ quasi-statically, in 10 steps. The iteration numbers exhibit "random" behavior, especially close to the jamming point. We plot both the average and the variance (as error bars) of the iteration number. We observe that the iteration number of each numerical method also follows a scaling law, which ranges from $10^3-10^4$ close to the jamming point ($\phi-\phic=10^{-6}$), to about hundreds when $\phi-\phic =10^{-2}$. The result shows that the preconditioned algorithm performs better, and also it is more robust to rounding error as the variance of the iteration number for the preconditioned algorithm is also smaller away from the jamming point.

In Table \ref{tab:methods}, we illustrate the performance of four optimization methods: L-BFGS, preconditioned L-BFGS, FR-CG,  and preconditioned FR-CG in three different situations given an over-jammed configuration with 4096 particles and volume fraction $\phi=0.8438$.
\begin{itemize}
\item Case 1, fixing the positions of the particles, and enlarging them slightly with ratio $1+\delta, \delta=5\times 10^{-5}$; 
\item Case 2, applying a small shear strain ($\gamma=10^{-4}$) to the configuration;
\item Case 3, exerting a small ($\mathcal{O}(10^{-4})$) random perturbation to the configuration. 
\end{itemize} 

The comparison of iteration number and computational time in Table \ref{tab:methods} clearly demonstrates that the preconditioner can improve the efficiency of the energy minimization methods. Also, it seems we can simply take the parameter $A=0$ \eqref{eq: precond matrix entry} in the preconditioner.

\newcommand{\tabincell}[2]{\begin{tabular}{@{}#1@{}}#2\end{tabular}}  
\begin{table}[H]
  \centering
    \begin{tabular}{|c|c|c|c|c|c|c|}
      \hline
      \multicolumn{2}{|c|}{Method}&L-BFGS&\tabincell{c}{P-L-BFGS\\($A=0$)}&\tabincell{c}{P-L-BFGS\\($A=3$)}&FR-CG&P-FR-CG\\
      \hline
      \multirow{2}*{Case 1}&n\_iter&2003&705&813&4574&1101\\
      \cline{2-7}
      &time/s&24.3&11.8&16.2&58.9&18.5\\
      \hline
      \multirow{2}*{Case 2}&n\_iter&2749&986&1351&7306&2013\\
      \cline{2-7}
      &time/s&29.2&15.8&27.1&90.1&33.5\\
      \hline
      \multirow{2}*{Case 3}&n\_iter&602&380&432&940&528\\
      \cline{2-7}
      &time/s&7.5&6.8&9.1&11.9&9.2\\
      \hline
    \end{tabular}
  \caption{Iteration number and computational time for L-BFGS,  P-L-BFGS, FR-CG, and P-FR-CG method for three different cases ($A$ is the parameter in \eqref{eq: precond matrix entry}). The granular system has 4096 particles and volume fraction $\phi=0.8438$. } \label{tab:methods}
\end{table}

\subsection{Shear Test}

In the shear test, we deform the over-jammed configuration with pure shear (stretching in x direction and compressing in y direction, while keeping the area unchanged). The deformation matrix is given by,
$$
D=\left(
    \begin{array}{cc}
      1+\gamma & 0 \\
      0 & \frac{1}{1+\gamma} \\
    \end{array}
  \right).
$$

\begin{figure}[H]
  \centering
   \includegraphics[width=0.9\linewidth]{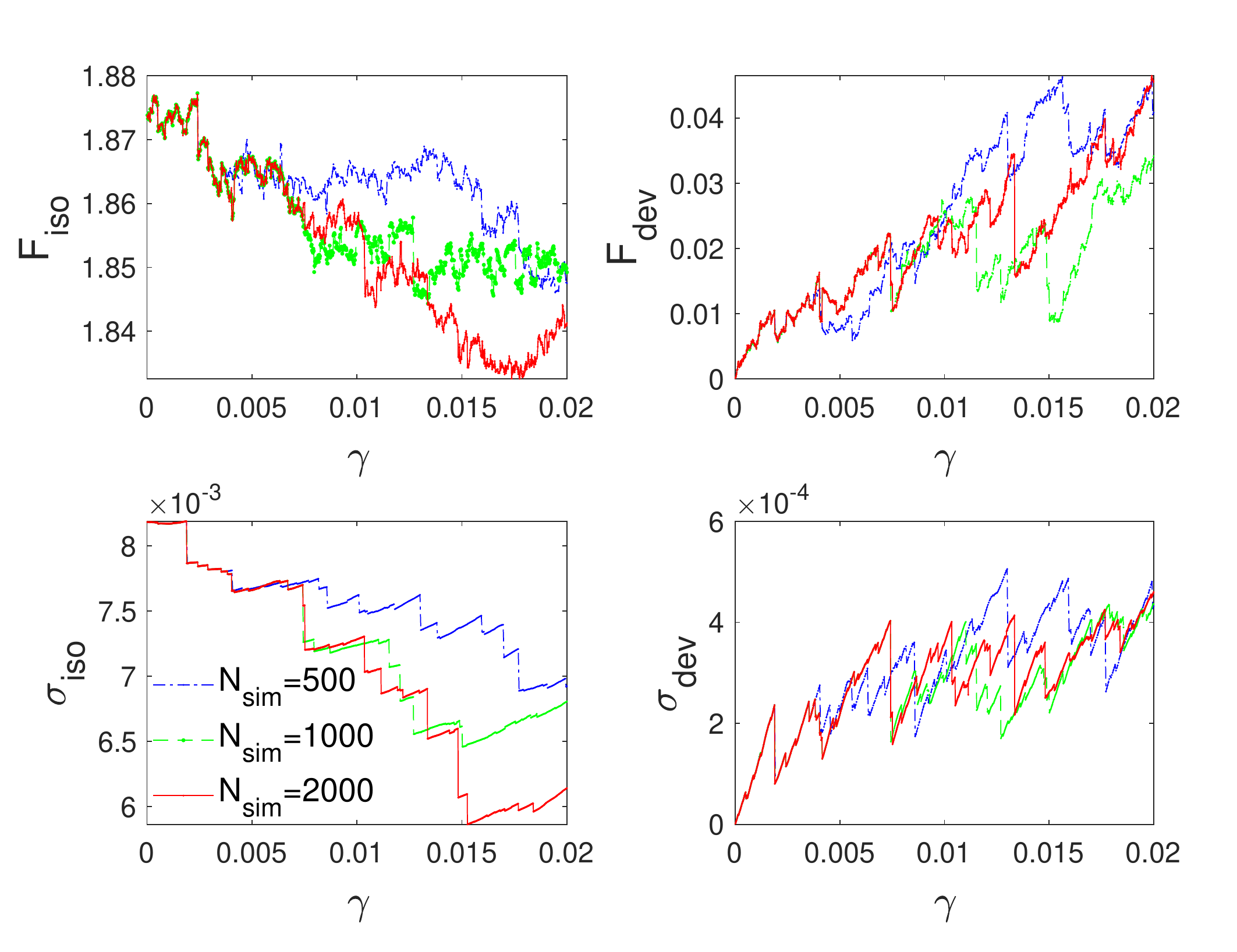}\\
  \caption{Stress ($\sigma$) vs. strain and fabric ($F$) vs. strain curve for pure shear, $\gamma$ is the shear strain. The subscripts iso and dev, represent the isotropic and deviatoric parts of these tensors (stress, fabric), respectively. Let $\lambda_1$ and $\lambda_2$ be the eigenvalues of those tensors, the isotropic part is defined as $(\lambda_1+\lambda_2)/2$ and the deviatoric part is $|\lambda_1 - \lambda_2|/2$} \label{fig:StressStrainCurvePureShear}
\end{figure}

The stress vs. strain and fabric vs. strain curve for the pure shear simulation is shown in Figure \ref{fig:StressStrainCurvePureShear}. Notice that we use increments of different sizes, $4\times 10^{-4}$, $2\times 10^{-4}$, and $10^{-4}$ to reach a final 2\% shear strain, we still obtain qualitatively (quantitatively for small strain) similar curves. 

\section{Conclusion}\label{sec:conclusion}
In this paper, we introduce the energy minimization techniques for the efficient simulation of dense athermal granular systems in two dimensions. Preconditioners are used to enhance the performance of the simulation. We carry out numerical experiments for some typical scenarios of granular simulation in order to validate the numerical methods, which include the jamming formation, scaling law phenomena close to the jamming point, and deformation of over-jammed states. Speedup of the preconditioned minimization methods are observed.

This work opens avenue for several possible improvements: First of all, we are going to extend the current work to three dimensions, which is physically more relevant to real applications, and still difficult for experimental studies \cite{Kou:2017, Zhang:2014} . 

Secondly, from the practical point of view, we will optimize the implementation of preconditioners using, for example, AMG \cite{Brandt:1985} and other types of state-of-the-art techniques, especially for the three dimensional case. 

Last but not least, we observe that the iteration number of our current preconditioned algorithms follows a scaling law behavior close to the jamming point, which is a fundamental bottleneck for the energy minimization approach. It remains open whether one can find a preconditioner which is robust close to the jamming point. Usually, an efficient preconditioner takes account of the long wavelength modes of the system. However, close to the phase transition point, more and more high frequency information might be needed. The development of numerical techniques in this direction relies on the understanding of the physical origin of the jamming transition.

\section*{Acknowledgments} 
We thank Jie Zhang and Zhaohui Jin for stimulating discussions on granular modeling and experiments. We thank Christoph Ortner and Mingjie Liao for stimulating discussions on numerical methods and their efficient implementations.

\medskip
\bibliographystyle{aims}
\bibliography{references}

\begin{thebibliography}{10}

\bibitem{Alrachid:2018}[10.5802/smai-jcm.29]
\newblock H. Alrachid, C. Ortner, and L. Mones,
\newblock Some remarks on preconditioning molecular dynamics, 
\newblock {\em SMAI Journal of Computational Mathematics}, \textbf{4} (2018).

\bibitem{Brandt:1985}
\newblock A. Brandt, S. McCoruick, and J. Huge,
\newblock Algebraic multigrid (amg) for sparse matrix equations, 
\newblock {\em Sparsity and its Applications}, \textbf{257} (1985).

\bibitem{Cuthill:1969}[10.1145/800195.805928]
\newblock E. Cuthill and J. McKee, 
\newblock Reducing the bandwidth of sparse symmetric matrices,
\newblock {\em In Proceedings of the 1969 24th national conference, ACM, New York, NY, USA}, (1969), 157--172

\bibitem{Edward:1924}
\newblock J. J. Edward,
\newblock On the determination of molecular fields.—II. From the equation of state of a gas,
\newblock {\em Proceedings of the Royal Society of London. Series A, Containing Papers of a Mathematical and Physical Character}, \textbf{106} (1924), 463--477

\bibitem{Fatih:2010}
\newblock G. Fatih, O. Durán and S. Luding,
\newblock Constitutive relations for the isotropic deformation of frictionless packings of polydisperse spheres,
\newblock {\em Comptes Rendus Mécanique}, \textbf{338} (2010), 570--586.

\bibitem{Fletcher:1964}[10.1093/comjnl/7.2.149]
\newblock R. Fletcher and C. M. Reeves,
\newblock Function minimization by conjugate gradients, 
\newblock {\em The computer journal}, \textbf{7} (1964), 149--154.

\bibitem{Hestenes:1952}
\newblock M. R. Hestenes and E. Stiefel, 
\newblock {Methods of conjugate gradients for solving linear systems},
\newblock {\em Journal of research of the National Bureau of Standards}, \textbf{49} (1952), 409--436.

\bibitem{Johnson:1987}
\newblock K. L. Johnson, 
\newblock {\em Contact Mechanics}, 
\newblock Cambridge university press, 1987.

\bibitem{Kou:2017}[10.1038/nature24062]
\newblock B. Kou, Y. Cao, J. Li, C. Xia, Z. Li, H. Dong, A. Zhang, J. Zhang, W. Kob and Y. Wang,
\newblock Granular materials flow like complex fluids, 
\newblock {\em Nature}, \textbf{551} (2017).

\bibitem{Krijgsman:2016}[10.1007/s40571-016-0105-8]
\newblock D.~Krijgsman and S.~Luding,
\newblock Simulating granular materials by energy minimization, 
\newblock {\em Computational Particle Mechanics}, \textbf{3} (2016), 463--475.

\bibitem{Luding:2008}
\newblock S. Luding,
\newblock Introduction to Discrete Element Methods: Basics of Contact Force Models, 
\newblock {\em European journal of environmental and civil engineering}, \textbf{12} (2008), 785--826.

\bibitem{Majmudar:2005}[10.1038/nature03805]
\newblock T. Majmudar and R. Behringer, 
\newblock Contact force measurements and stress-induced anisotropy in granular materials, 
\newblock {\em Nature}, \textbf{435} (2005), 1079--1082.

\bibitem{Mones:2018}[10.1038/s41598-018-32105-x]
\newblock L. Mones, C. Ortner and G. Csányi, 
\newblock Preconditioners for the geometry optimisation and saddle point search of molecular systems, 
\newblock {\em Scientific Reports}, \textbf{8} (2018).

\bibitem{Nocedal:1980}[10.2307/2006193]
\newblock J. Nocedal, 
\newblock Updating quasi-Newton matrices with limited storage,
\newblock {\em Mathematics of computation}, \textbf{35} (1980), 773--782

\bibitem{Nocedal:2006}
\newblock J. Nocedal and S. J. Wright,
\newblock {\em Numerical Optimization}, 
\newblock 2$^{nd}$ edition, Springer Science \& Business Media, New York, USA, 2006.

\bibitem{Ogden:1997}
\newblock R.W. Ogden, 
\newblock {\em Non-linear Elastic Deformations}, 
\newblock Courier Corporation, 1997.

\bibitem{O'Hern:2003}[doi.org/10.1103/PhysRevE.68.011306]
C.~S. O'Hern, L.~E. Silbert, A.~J. Liu, and S.~R. Nagel, 
\newblock Jamming at zero temperature and zero applied stress: The epitome of disorder,
\newblock {\em Phys. Rev. E}, \textbf{68} (2003), 011306.

\bibitem{Packwood:2016}[10.1063/1.4947024]
D. Packwood, J. Kermode, L. Mones, N. Bernstein, J. Woolley, N. Gould, C. Ortner, and G. Cs{\'a}nyi, 
\newblock A universal preconditioner for simulating condensed phase materials, 
\newblock {\em The Journal of Chemical Physics}, \textbf{144} (2016), 164109.

\bibitem{Pytlak:2008}
R. Pytlak,
\newblock {\em Conjugate gradient algorithms in nonconvex optimization, Vol~89},
\newblock Springer Science \& Business Media, 2008.

\bibitem{Speedy:1999}[10.1063/1.478337]
R.~J. Speedy,
\newblock Glass transition in hard disc mixtures,
\newblock {\em The Journal of chemical physics}, \textbf{110} (1999), 4559--4565.

\bibitem{Stukowski:2009}[10.1088/0965-0393/18/1/015012]
\newblock A. Stukowski,
\newblock Visualization and analysis of atomistic simulation data with {OVITO}{\textendash}the Open Visualization Tool,
\newblock {\em Modelling and Simulation in Materials Science and Engineering}, textbf{18} (2009), 015012.

\bibitem{Wathen:2015}[10.1017/S0962492915000021]
\newblock A. J. Wathen,
\newblock Preconditioning,
\newblock {\em Acta Numerica}, \textbf{24} (2015), 329--376.

\bibitem{Zhang:2005}[10.1103/PhysRevE.72.011301]
H. Zhang and H.~A. Makse, 
\newblock Jamming transition in emulsions and granular materials, 
\newblock {\em Phys. Rev. E}, \textbf{72} (2005), 011301.

\bibitem{Zhang:2014}[10.1103/PhysRevE.89.012203]
L. Zhang, Y. Wang and J. Zhang.
\newblock Force-chain distributions in granular systems.
\newblock {\em Phys. Rev. E}, \textbf{89} (2014), 012203.

\end{thebibliography}

\medskip
Received xxxx 20xx; revised xxxx 20xx.
\medskip

\end{document}